\documentclass[dvips,12pt]{article}
\usepackage{amsmath}
\usepackage{amssymb}
\def\L{{\Omega_\Lambda}}    \def\M{{\Omega_m}}       \def\K{{\Omega_k}}

\begin{document}

\newpage

\setcounter{page}{0}

\title{ The function q(z) as a consistency check for cosmological parameters}

\vspace{1cm}

\author{ Maria Luiza Bedran \thanks{On leave from Universidade Federal do Rio de Janeiro.}\thanks{ Departamento de F\'{\i}sica - ICE - Universidade Federal de Juiz de Fora, MG, Brazil. Email:bedran@fisica.ufjf.br} }

\vspace{1cm}

\maketitle

\begin{abstract}

In the Friedmann cosmology the deceleration of the expansion $q$ plays a fundamental role. We derive the deceleration as a function of redshift $q(z)$ in two scenarios: $\Lambda$CDM model and modified Chaplygin gas (MCG) model, without assuming spatial flatness. The expression relating the transition redshift from decelerated to accelerated expansion $z_t$ to the cosmological parameters is obtained; it is seen that the curvature parameter does not appear in this expression. Of course the numerical value of $z_t$ depends on $\K$, since $\L+\M+\K=1$. The exact function $q(z)$ allows the calculation of an inflection point at $z\approx0.1$ for the flat $\Lambda$CDM model. This inflection point is visible in Fig.5 of the work of Daly and Djorgovski, Ap.J. 612,652 (2004), where the deceleration was plotted directly from observational data. We also consider the possibility of a small contribution of the curvature, namely, $\K_0=-0.1\L_0$, and calculate the transition redshift to be $z_t=0.56$, which falls inside the interval $0.46\pm0.13$ obtained by Riess et al., Ap.J. 607, 665 (2004). The calculated transition redshift for the MCG model is considerably lower $z_t\approx0.2$, but the observational data do not exclude a pure Chaplygin gas.

\end{abstract}



\newpage

\section{Introduction}
During the last decade, the observation of type IA supernovae
(SNeIA) and the cosmic backgroud radiation (CBR) permitted the
determination of the cosmological parameters with ever increasing
precision. The reported results of the five year analysis of WMAP
\cite{Komatsu} are $\L_0=0.721 \pm 0.015$ and $\M_0=0.279 \pm
0.015$~; these values were obtained assuming a flat geometry
($\K_0=0$). From measured luminosity distances to SNeIA, Riess et
al.\cite{Riess} determined the redshift of the transition from
decelerated to accerelerated expansion to be $z_t=0.46 \pm 0.13$~;
this value was obtained assuming a linear expansion for the
deceleration parameter, thay is,~ $q(z)=q_0+q_1 z$.\\
 In this paper we derive the exact function for $q(z)$ in two
 different scenarios: cold dark matter + cosmological constant
 ($\Lambda CDM$) and modified Chaplygin gas (MCG). This derivation
 was performed with the minimal assumptions: Einstein's general
 relativity is valid and the universe is homogeneous and
 isotropic, that is, the cosmological metric is the
 Friedmann-Robertson-Walker (FRW) one. {\it No assumption was made
 about the spatial curvature of the universe}.\\
 We obtained the  functions $q(z)$ for the two models in terms of
 the cosmological  parameters ~ $\L$,~$\M$,~$\K=1-\L-\M$~. These 
 functions are simple even in the non-flat cases and can be compared 
 with the observational data. The calculation 
 of the derivatives of $q(z)$ is illuminating and show an inflection 
 point in the $\Lambda$CDM model. Then we determined the
 transition redshift $z_t$ where $q(z_t)=0$. For the $\Lambda$CDM
 model we obtain the condition $1+z_t =( 2\L_0 / \M_0)^{1/3}$~,
 which is not  satisfied by the reported cosmological parameters of the previous
 paragraph. Considering the possibility of non-flat geometry, ~$\K_0 <0$,
 the calculated transition redshift is consistent with $z_t=0.46 \pm 0.13$
 obtained by Riess et al.\cite{Riess}.
 For the MCG model, with equation of state $p=B \rho  -A\rho^{-\alpha}$,
 the measured parameters are consistent with the theoretical calculation 
 if we  choose $B=0$ and $\alpha \approx 0.4$.\\
 Future measurements of luminosity distances to high redshift supernovae 
 will allow a better determination of the transition
 redshift $z_t$. We believe that the exact function $q(z)$ can 
 help in this determination and provide another means of checking the energy
 parameters ~ $\L$ and $\M$.

\section{$\Lambda$CDM model}
Let's assume that the universe is described by the FRW metric and the energy content is a pressureless fluid and a cosmological constant. The first Friedmann equation for the scale factor $a(t)$ reads

\begin{equation}
H^2=\frac{\dot a ^2}{a^2}=\frac{8\pi G\rho}{3}+\frac{\Lambda}{3}-\frac{k}{a^2}
\label{Friedmann}
\end{equation}

\noindent which can be written as

\begin{equation}
\M+\L+\K=1                            \label{Friedmann1}
\end{equation}

\noindent with the definitions

$$ \M = \frac{8\pi G \rho}{3H^2}~~~~~~~~\L=\frac{\Lambda}{3H^2} ~~~~~~~~\K=\frac{-k}{a^2H^2}$$

The Bianchi identity for a pressureless fluid $$\dot{\rho}+3\frac{\dot a}{a} \rho = 0 $$ can be integrated to give

\begin{equation}
\rho a^3 = \rho_0 a_0^3                          \label{Bianchi}
\end{equation}

\noindent where the subscript zero denotes present values of the quantities. Inserting eq.(\ref{Bianchi}) into eq.(\ref{Friedmann}) and writing $-k$ as 

$$ -k = H^2 a^2 \K = H_0^2 a_0^2 (1-\M_0 -\L_0) $$

\noindent we obtain

\begin{equation}
{\dot a}^2= H_0^2 a_0^2 \left[ \M_0~\frac{a_0}{a} + \L_0~\frac{a^2}{a_0^2} + \K_0 \right]
\label{eq.4}
\end{equation}

For observational reasons it is convenient to use the redshift $z$ instead of the scale factor $a(t)$. As $$ 1+z=\frac{a_0}{a}$$ eq.(4) gives 

\begin{equation}
\dot a^2 = \frac{H_0^2 a_0^2}{(1+z)^2}~Q(z)= H_0^2 a^2(z) Q(z)      \label{eq.5}
\end{equation}

\noindent where

\begin{equation}
Q(z)=\M_0~(1+z)^3 + \K_0~(1+z)^2 + \L_0                \label{Q(z)}
\end{equation}
\vspace{.5cm}

Using eq.(5) we can relate $\frac{da}{dz}$ with $\dot a =\frac{da}{dt}$:

\begin{equation}
\frac{da}{dz}=-\frac{a_0}{(1+z)^2}=-\frac{a(z)}{1+z}=-\frac{1}{H_0 (1+z) \sqrt{Q(z)}}~\frac{da}{dt}                                                \label{da/dz}
\end{equation}

\noindent from which we infer the relation between $dt$ and $dz$:

\begin{equation}
\frac{dz}{dt}= - H_0 (1+z) \sqrt{Q(z)}                \label{dz/dt}
\end{equation}

Now, using eq.(\ref{da/dz}), we can calculate the second derivative of $a(t)$ in terms of the redshift $z$:

\begin{equation}
\ddot a = \frac{dz}{dt}\frac{d\dot a}{dz}= \frac{dz}{dt}\frac{d}{dz}\left[-H_0 F(z) \frac{da}{dz}\right]= -H_0 \frac{dz}{dt}\left[\frac{dF}{dz}\frac{da}{dz} + F(z)\frac{d^2a}{dz^2}\right]                     \label{eq.9}
\end{equation}

\noindent where

\begin{equation}
F(z)=(1+z)\sqrt{Q(z)}=\left[ \M_0(1+z)^5 + \K_0(1+z)^4 + \L_0(1+z)^2 \right]^{1/2}    \label{F(z)}
\end{equation}
\vspace{.5cm}

The derivatives in eq.(\ref{eq.9}) can be easily calculated, yielding an expression for $\ddot a$ in terms of $z$ and the cosmological parameters. It is interesting to consider the redshift $z_t$ of the transition from decelerated to accelerated expansion. At the moment when $\ddot a=0$, the above equations yield

\begin{equation}
(1+z_t)^3 = \frac{2\L_0}{\M_0}                 \label{zt}
\end{equation}

The transition redshift $z_t$ can be obtained from measured luminosity distances of type IA supernovae, as well as the cosmological parameters $\L_0$ and $\M_0$. Note that no assumption was made for the curvature parameter $\K_0$; the curvature can be computed from $\K_0=1-\M_0-\L_0$. Recent values of the cosmological parameters reported by the WMAP team \cite{Komatsu} are $\L_0 = 0.721\pm 0.015$ and $\M_0 = 0.279\pm0.015$. When inserted into eq.(\ref{zt}) these values yield $z_t=0.73\pm0.04$, which is far beyond the value $z_t= 0.46\pm0.13$ obtained by Riess et al.\cite{Riess} with SNeIA data, considering a linear expansion for the deceleration parameter, that is,~ $q(z)=q_0+q_1 z$~. It is worth to mention that the $\L,\M$ parameters of the WMAP were obtained from the CBR observations assuming $\K=0$. For $\K=0, \M=1-\L$, the consistency condition (\ref{zt}) with $z_t=0.46$ would require $\L\approx0.6$. On the other hand, in ref.\cite{Leibundgut} Leibundgut and Blondin claim that {\it the most recent determination of the cosmological parameters by the HZT favour values that are rather different from a flat universe solution}. Using the best values of ref.\cite{Leibundgut}~$\L_0=1.25~,~\M_0=0.7$~in eq.(\ref{zt}), we obtain $z_t=0.53$, which is consistent with $0.46\pm0.13$. For a review of the limitations in the accuracy of supernovae data see ref.\cite{Leibundgut2}.\\

Let's derive the function $q(z)$ exactly. The definition of the deceleration parameter $q$ is

\begin{equation}
q = - \frac{ \ddot a}{H^2 a}             \label{q}
\end{equation}

Now eq.(9) reads

\begin{equation}
\ddot a= -H_0 \frac{dz}{dt}\left[-\frac{a_0}{(1+z)^2}\frac{dF}{dz}+ \frac{2a_0}{(1+z)^3} F(z) \right]                                     \label{ddota}
\end{equation}         

Inserting eqs.(\ref{dz/dt}),(\ref{F(z)}) and (\ref{ddota}) into eq.(\ref{q}) we are led to:

\begin{equation}
q(z)=-\frac{H_0^2}{H^2 (1+z)^2}\left[-\frac{\M_0}{2}(1+z)^5 + \L_0(1+z)^2\right]
\end{equation}

\noindent which, upon sustitution of $H^2=H_0^2Q(z)$, reads

\begin{equation}
q(z)= \left[\frac{\M_0}{2}(1+z)^3-\L_0\right]\left[\M_0(1+z)^3+\K_0(1+z)^2+\L_0\right]^{-1}
\label{q(z)}
\end{equation}  \vspace{.5cm}

\noindent For $z>>1$ we see that $q(z)=0.5$ as expected. It is interesting to write some particular cases of the function $q(z)$.\\

i)For $\L_0=0$, which implies $\K_0=1-\M_0$, we have $$q(z)=\frac{\M_0(1+z)}{2(1+z\M_0)}\geq 0.$$ \\
In the absence of a cosmological constant the expansion is decelerated at any redshift.\\

ii)For $\K_0=0~\Rightarrow ~ \M_0=1-\L_0$,

 $$q_{flat}(z)=\left[\frac{(1-\L_0)(1+z)^3}{2} - \L_0\right]\left[(1-\L_0)(1+z)^3+\L_0\right]^{-1},$$
  which implies $q<0$~ for~ $z<\left(\frac{2\L_0}{1-\L_0}\right)^{1/3} -1 $.\\
  
  For a flat universe with a cosmological constant, the expansion is decelerated at large $z$ and becomes accelerated at $z_t$; the expansion cannot decelerate again.\\

Let's calculate the derivatives of the function $q(z)$ (eq.\ref{q(z)}):

\begin{equation}
q'(z)=Q^{-2}(z)~(1+z)\left[\frac{\M_0\K_0}{2}(1+z)^3 + \frac{9}{2}\M_0\L_0(1+z) +2\L_0\K_0\right]               \label{q'}
\end{equation}

\begin{align} 
-~q''(z)= &Q^{-3}(z)\left\{\M_0(1+z)\left[\M_0\K_0(1+z)^5 + 18\M_0\L_0(1+z)^3 \right.\right. \nonumber \\
  &  \left.  + 17\L_0\K_0(1+z)^2 - 9(\L_0)^2\right] \nonumber \\
  &\left.  + 2\L_0\K_0\left[3\K_0(1+z)^2 - \L_0\right] \right\}     \label{q''}
\end{align}

In the flat case $\K_0=0$ they reduce to:

$$q'_{flat}(z)=\frac{9}{2}\M_0\L_0(1+z)^2Q^{-2}(z) ~~>0,~~\forall z$$

$$-q''_{flat}(z)=9\M_0\L_0(1+z)[2\M_0(1+z)^3-\L_0]Q^{-3}(z)$$\\

Let's compute the inflection point of $q_{flat}(z)$ with the cosmological parameters $\L_0=0.72$ and $\M_0=0.28$; we obtain $z_i=0.09$, with $q''>0$ for $z<z_i$ and $q''<0$ for $z>z_i$.\\

In ref.\cite{Daly} Daly and Djorgovski performed a direct determination of the kinematics of the universe using measured distances to SNeIA at redshifts ranging from $0.01$ up to $1.8$. The plot of $q(z)$ in their Fig.5 shows an inflection point at $z\approx0.1$, in good agreement with our calculation, but they obtained $q_0=q(z=0)=-0.35\pm0.15$ and $z_t\approx0.4$~, while our calculation yields $q_0=-0.58$ and $z_t\approx0.7$ for the flat case.\\

Now suppose that the spatial curvature is not zero and that $\K$ is negative and one order of magnitude smaller than $\L$, that is, $\K_0=-0.1 \L_0 $. With the values $\L_0=0.70$,~$\K_0=-0.07$ ~and $\M_0=0.37$,~eq.(\ref{q(z)}) gives $q_0=-0.51$ and $z_t=0.56$~. In this case $q(z)$ has a maximum at $z\approx8.5$,~where $q=0.51$,~before attaining the asymptotic value $0.5$. An extensive discussion of the degeneracies between dark energy and cosmic parameters was done in ref.\cite{Hlozek}.\\

If we expand eq.(\ref{q(z)}) in a Taylor series around $z=0$, we obtain:

  $$q(z)\approx \left(\frac{\M_0}{2}-\L_0\right)+\left(2\K_0\L_0+\frac{9\M_0\L_0}{2}+\frac{\M_0\K_0}{2}\right)z + O(z^2)$$\\
  
  For the flat case, $\K_0=0,~\L_0=0.72,~\M_0=0.28$~ , this yields 
  
  $$q_{flat}(z)\approx -0.58+0.91~z+0.14~z^2~\Rightarrow ~q_{flat}(z=0.46)\approx-0.13$$    
  which shows that $z=0.46$ cannot be the transition redshift in this model. \\
  
\vfill
  
\section{Modified Chaplygin gas (MCG) model without cosmological constant}
  
  Let's change the energy content of the universe to a fluid that behaves like a perfect fluid of non-zero pressure at early times, like a pressureless fluid at intermediate times and like dark energy at present. The generalized Chaplygin gas model has been proposed as a source term in Einstein's field equations in order to unify the concepts of cold dark matter and dark energy \cite{Kamenshchik,Bilic,Bento}. Constraints on the parameters of the modified Chaplygin gas from recent observations were obtained in ref.\cite{Lu}. \\
  
  The equation of state of the MCG is given by
  
  \begin{equation}
p = ~B\rho - \frac{A}{{\rho}^{\alpha}}           \label{eq.state} \,,
\end{equation}

\noindent where $A$,$B$ and $\alpha$ are non-negative constants. For $B=0$ we have the pure Chaplygin gas and for $A=0$ a perfect fluid. Now the Bianchi identity 

\begin{equation}
3\,{\frac {\dot a} a}\,(p + \rho) + \dot {\rho} =0 \,
\end{equation}

\noindent yields after integration (see ref.\cite{Bedran} for details):

\begin{equation}
\rho (a) = \rho _0 \left[ {\Omega _X  + \left( {1 - \Omega _X } \right)\left( {\frac{{a_0 }}{a}} \right)^{3R} } \right]^{{1 \mathord{\left/
 {\vphantom {1 {\left( {\alpha  + 1} \right)}}} \right.
 \kern-\nulldelimiterspace} {\left( {\alpha  + 1} \right)}}}  \,,
\end{equation}

\noindent where $\rho_0$ is the present energy density, $R=(B+1)(\alpha +1)$, and we define the dimensionless parameter

\begin{equation}
\Omega_X = \frac{A}{(B+1)\,\rho_0^{\alpha+1}} \,.         \label{OmegaX}
\end{equation}

The same analysis of the previous section can be done, but now the function $Q(z)$ changes to:

\begin{equation}
Q(z)= \left\{\left[\Omega_X + (1-\Omega_X)(1+z)^{3R}\right]^{1/(1+\alpha)}+\K_0(1+z)^2\right\}^{1/2},
\end{equation}

\noindent leading to the following expression for the deceleration parameter:

\begin{equation}
q(z)=\frac{Q^{-3/2}(z)}{2} \left[\Omega_X+(1-\Omega_X)(1+z)^{3R}\right]^{-\alpha/(1+\alpha)}\left\{(1-\Omega_X)(3B+1)(1+z)^{3R} - 2\Omega_X\right\}.                 \label{q(z)MCG}
\end{equation}

The transition redshift for the MCG model is given by:

\begin{equation}
(1+z_t)^{3R}=\frac{2\Omega_X}{(3B+1)(1-\Omega_X)}.   \label{ztMCG}
\end{equation}

The dimensionless parameter $\Omega_X$ represents the fraction of dark energy in the content of the universe, thus taking the value $\Omega_X=0.72$. If we chose $B=1/3$ in order to describe the evolution of the universe since the radiation era, and consider $0 <\alpha < 0.5$, which is required from thermodynamical considerations (see ref.\cite{Santos}), we find $z_t \approx 0.2$. If we consider a pure Chaplygin gas $(B=0)$ and  $\Omega_X=0.72$, the value $z_t=0.46$ can be achieved if $\alpha \approx 0.4$. This result is compatible with the analysis done in ref.\cite{Makler} for a Chaplygin gas in the flat ($k=0$) case.

\section{Discussion} 

We have derived expressions for the deceleration parameter $q$ as functions of the redshift $z$ and the observable parameters $\M~,~\L,~\K$ in two different cosmological models: $\Lambda$CDM and Modified Chaplygin Gas (MCG).\\

 The only assumptions were:\\
i)The geometry of the universe is described by the FRW metric. {\it No assumption was made for the scalar curvature.}\\
ii)The field equations are those of Einstein's theory of relativity.\\

The assupmtions that the universe is described by the FRW metric and the $\Lambda$CDM model lead to a relation between the cosmological parameters $\L_0$~,~$\M_0$ and the transition redshift $z_t$ (see eq.(\ref{zt}) from decelerated to accelerated expansion that is not satisfied by the reported values of these parameters in \cite{Komatsu,Riess}, to wit, $\L_0 = 0.721\pm 0.015$,~ $\M_0 = 0.279\pm0.015$ and $z_t=0.46 \pm 0.13$. The parameters determined in \cite{Komatsu,Riess} were obtained with the assumption that $\K_0=0$. So it seems that the  evidences of the spatial flatness of the universe are not so strong. In ref.\cite{Leibundgut} the authors found a degeneracy in the $\L vs. \M$ plane along a line corresponding to $\L - 1.4\M=0.35\pm0.14$ and showed that the flat universe model is not within the 68\% likelihood area. Considering the best values determined in \cite{Leibundgut}, $\L_0=1.25,~\M_0=0.7~\Rightarrow \K_0=-0.95$, the transition redshift becomes $z_t=0.53$, which falls inside the interval $0.33<z_t<0.59$ obtained by Riess et al. \cite{Riess}. \\
The exact expression for $q(z)$ for the $\Lambda$CDM model also allowed the calculation of an  inflection point $q''(z_i)=0$. In the flat case it gives $z_i=0.09$. The inflection point is visible in Fig.5 of ref.\cite{Daly} at $z\approx0.1$.\\
Our calculations with the MCG model showed that the pure Chaplygin gas is compatible with the observational parameters, even in the flat case. \\
We conclude by saying that exact functions derived with the minimal assumptions are always useful, and we hope that the function $q(z)$ can help in the analysis of observational data.

\end{document}